\documentclass[a4paper,aps,onecolumn,nofootinbib,preprintnumbers,superscriptaddress,amsmath,amssymb,amsfonts]{revtex4}
\usepackage{graphicx}
\usepackage{bm,natbib}
\usepackage{amsmath}
\usepackage[english]{babel}
\usepackage[T1]{fontenc}
\usepackage{amssymb}
\usepackage{amsfonts}
\usepackage{epsfig}
\usepackage{colordvi}
\usepackage{psfrag}
\usepackage{color}
\usepackage{ mathrsfs }
\newcommand{\Lagr}{\mathcal{L}}

\newcommand{\Gi}{\mathcal{G}}
\newcommand{\al}{\alpha}
\newcommand{\be}{\beta}

\newcommand{\La}{\mathscr{L}}
\usepackage{dcolumn}
\usepackage{multirow}
\usepackage{hyperref}
\usepackage{epstopdf}
\usepackage{bm}
\usepackage{rotating}
\usepackage{lscape}
\usepackage{times}
\usepackage{comment}

\hypersetup{
  colorlinks=true,        
  linkcolor=blue,         
  citecolor=magenta,      
}

\begin{document}
\title{Minisuperspace Quantum Cosmology in Metric and Affine Theories of Gravity}

\author{Salvatore Capozziello}
\email{capozziello@na.infn.it}
\affiliation{Scuola Superiore Meridionale, Largo San Marcellino 10, I-80138, Naples, Italy.}
\affiliation{Department of Physics ``E. Pancini'', University of Naples ``Federico II'', Naples, Italy.}
\affiliation{INFN Sez. di Napoli, Compl. Univ. di Monte S. Angelo, Edificio G, Via Cinthia, I-80126, Naples, Italy.}

\author{Francesco Bajardi}
\email{francesco.bajardi@unina.it}
\affiliation{Scuola Superiore Meridionale, Largo San Marcellino 10, I-80138, Naples, Italy.}
\affiliation{Department of Physics ``E. Pancini'', University of Naples ``Federico II'', Naples, Italy.}
\affiliation{INFN Sez. di Napoli, Compl. Univ. di Monte S. Angelo, Edificio G, Via Cinthia, I-80126, Naples, Italy.}

\date{\today}

\begin{abstract}
Minisuperpace Quantum Cosmology is an approach by which it is possible to infer initial conditions for dynamical systems which can suitably represent \textit{observable} and \textit{non-observable} universes. Here we discuss theories of gravity which, from various points of view, extend Einstein's General Relativity.  Specifically,   the Hamiltonian formalism for $f(R)$, $f(T)$ and $f(\Gi)$ gravity, with $R$, $T$, and $\Gi$ being the curvature,   torsion  and  Gauss--Bonnet scalars, respectively, is developed starting from the Arnowitt-Deser-Misner approach. The Minisuperspace Quantum Cosmology is  derived for all these models and cosmological solutions are  obtained thanks to the existence of Noether symmetries. The Hartle criterion allows the interpretation of solutions in view of observable universes.
\end{abstract}

\maketitle

\section{Introduction}
\label{introd}
The Arnowitt-Deser-Misner (ADM) formalism has been developed in 1962 to the purpose of solving issues occurring in the attempt to merge the formalism of General Relativity (GR) with Quantum Mechanics \cite{Arnowitt:1962hi}. By means of a 3+1 decomposition of the metric, it is possible to get a gravitational Hamiltonian and find canonical quantization rules leading to a Schr\"odinger-like equation, dubbed the \emph{Wheeler-De Witt} (WDW) equation, firstly obtained by J. A. Wheeler and B. DeWitt in 1967 \cite{DeWitt:1967yk, DeWitt:1967ub, Wheeler:1957mu}. Nowadays, the ADM formalism is not considered as the ultimate candidate to solve the quantization problem    of GR, both because it does not account for a full theory of Quantum Gravity and because it implies an infinite-dimensional superspace which cannot be easily handled. However, the restriction of the problem to cosmology turns out to be useful for several reasons. On the one hand, the configuration superspace can be reduced to a finite-dimensional minisuperspace, where the WDW equation can be  analytically solved. On the other hand, Quantum Cosmology can provide information regarding the very early stages of the universe evolution by means of the so called \emph{Wave Function of the Universe}, which is the solution of the  WDW equation. Clearly, this wave function cannot be interpreted as a straightforward probability amplitude like in  Quantum Mechanics, due to the lack of a Hilbert space and a definite-positive inner product in gravitational theory. Moreover, the probabilistic meaning, based on many copies of the same system, cannot be applied in the standard Copenhagen interpretation of Quantum Mechanics.

For these reasons, its meaning is still unclear, though different interpretations have been proposed. For instance, it can be thought as an indication of the probability  for the quantum system to evolve towards our classical universe~\cite{Vilenkin:1988yd, Hawking:1983hj}. Another interpretation, related to the enucleation from nothing, was provided by Vilenkin in \cite{Vilenkin:1982de, Vilenkin:1984wp}.  Furthermore, according to the so called \emph{Many World Interpretation} \cite{Everett}, the wave function is supposed to come from quantum measurements that are simultaneously realized in different universes, without showing any collapse, as in standard Quantum Mechanics~\cite{Bousso:2011up}. In 1983 J.~B.~Hartle proposed to consider the wave function trend: when it is    oscillating, variables are correlated and then  it is possible to recover observable universes. The  analogy comes from  the wave function interpretation of non-relativistic Quantum Mechanics. This interpretation  is the so called \emph{Hartle criterion}~\cite{Hartle:1983ai}. Interestingly, using this criterion, in the semiclassical limit one can recast the wave function as $\psi = e^{i m_P^2 S}$, with $S$ being the action and $m_P$ the Planck mass. As a consequence, classical trajectories can be straightforwardly found by means of the Hamilton-Jacobi equation. 

The formalism of Quantum Cosmology has been successfully applied to GR in the attempt to solve open issues related to the first phases of the Universe. However, in the high-energy regime, GR exhibits several other shortcomings. For this reason, many alternative models have been proposed over the years, such as \emph{Kaluza-Klein Theory} \cite{Klein:1926tv, Kaluza:1921tu, Han:1998sg, Servant:2002aq, Duff:1986hr}, \emph{String Theory} \cite{Green:1987sp, Polchinski:1998rq, Seiberg:1999vs, Witten:1995ex, Friedan:1985ge}, \emph{Non-Local Gravity} \cite{Simon:1990ic, Modesto:2013jea, Koshelev:2016xqb, Calcagni:2013vra, Capozziello:2022lic}, \emph{Loop Quantum Gravity} \cite{Ashtekar:2011ni, Rovelli:1997yv, Rovelli:1989za, Meissner:2004ju, Bojowald:2005epg}, \emph{etc}. 

All these theories try to address small-scales shortcoming suffered by GR, and most of them recover Quantum Cosmology as a limit \cite{Bojowald:2015iga}.

However, matching the quantum formalism is not the only reason to consider GR alternatives, since many problems also occur at cosmological and astrophysical scales. More precisely, though GR predictions have been successfully confirmed by many experiments and observations, the theory also exhibits  incompatibilities  at several scales of energy \cite{Bull:2015stt, Clifton:2011jh}. Two of the most controversial problems are related to the existence of Dark Energy and Dark Matter. The former should represent most of the universe content and should be made by a never detected fluid with negative pressure, introduced to address the today observed  accelerating expansion; the latter was firstly introduced to fit the galaxy rotation curves. For a detailed discussion on puzzles and shortcomings in  Einstein's gravity see \emph{e.g.} \cite{Faraoni:2010pgm}. 

Due to the need of addressing incompatibilities between GR and observations, several modified/extended theories of gravity arose in the last decades. Within the landscape of modified models, some theories extend the Hilbert-Einstein action by including functions of second-order curvature invariants \cite{Nojiri:2005jg, Bueno:2016ypa}, some relax GR assumptions such as the Lorentz Invariance \cite{Colladay:1998fq, Collins:2004bp, Horava:2009uw}, the Equivalence Principle \cite{Hui:2009kc, Damour:2010rp}, or the metric-compatible connection \cite{Olmo:2011uz}. 

One of the most straightforward extension is the so called $f(R)$ gravity, whose action contains a function of the scalar curvature. For  detailed reviews see \cite{Capozziello:2011et, Nojiri:2017ncd}. By extending the gravitational action, it is possible to find out explanations for open questions at astrophysical and cosmological level. For instance,  $f(R)$ gravity, in the post-Newtonian limit,  is capable of fitting the galaxy rotation curve without introducing dark matter \cite{Sanders:2006sz, Capozziello:2006ph}, the cosmological  accelerating expansion without any dark energy \cite{Copeland:2006wr, Capozziello:2002}, or the mass-radius diagram of neutron stars without exotic equations of state \cite{Astashenok, Oikonomou}. However, to date, no $f(R)$ model can account for the ultimate candidate capable of solving all the problems exhibited by GR simultaneously at any scale.

Furthermore, the Hilbert-Einstein action can be  extended by considering also other second-order curvature invariants. An example is $f(R, R^{\mu \nu} R_{\mu \nu}, R^{\mu \nu \rho \sigma} R_{\mu \nu \rho \sigma})$ gravity, with $R_{\mu \nu}$ and $R_{\mu \nu \rho \sigma}$ being the Ricci and the Riemann tensors, respectively. These additional terms naturally arise from the one-loop effective action of GR, with the consequence that the renormalization procedure can be pursued only when higher-order invariants are included, so that UV divergences can be avoided \cite{Stelle:1976gc, Adams:1990pn, Cotsakis:2006zn, Amendola:1993bg}. Among all possible choices, there is only a particular combination of curvature invariants giving a topological surface in four dimensions. It is $\Gi = R^2 - 4 R^{\mu \nu} R_{\mu \nu} + R^{\mu \nu \rho \sigma} R_{\mu \nu \rho \sigma}$, the \emph{Gauss--Bonnet invariant}. As a consequence, the integration of $\Gi$ over the given manifold provides the Euler characteristic, namely a topological invariant. In four dimensions or less, $\Gi$ vanishes identically, while in more than four dimensions it provides non-trivial contributions to the field equations. Due to the impossibility of dealing with the linear term $\Gi$ in four dimensions, usually a function of the Gauss--Bonnet term, that is $f(\Gi)$, is considered into the action. In this way, the corresponding field equations can contribute to the dynamics, since $f(\Gi)$ starts being trivial in three dimensions or less. Therefore,  the gravitational action can be extended by considering either the function $f(\Gi)$ or, in general, $f(R, \Gi)$. Both $f(R,\Gi)$ and $f(\Gi)$ models exhibit several interesting features and provide some explanations for unsolved problems of GR at large scales \cite{Nojiri:2005vv, Cvetic:2001bk, Bajardi:2020osh, Glavan:2019inb}. Being a topological surface, the Gauss--Bonnet term can reduce dynamics and provide analytic solutions for the field equations; moreover, it naturally emerges in gauge theories of gravity such as Lovelock, Born-Infeld or Chern-Simons gravity \cite{Bajardi:2021hya, Zanelli:2005sa, Mardones:1990qc, Ferraro:2008ey}.

Another class of alternatives to GR is represented by those models relaxing the assumption of torsionless and metric-compatible connections. In particular, it is possible to show that the most general affine connection is made of three different contributions, respectively related to curvature, torsion and non-metricity. Torsion in the space-time occurs when the connection exhibits an anti-symmetric part, that is when $\Gamma^\alpha_{\mu \nu} \neq \Gamma^\alpha_{\nu \mu}$. Non-metricity occurs when $\nabla_\alpha g_{\mu \nu} \neq 0$, with $\nabla$ being the covariant derivative. Therefore, gravity can be described by means of torsion, curvature and/or non-metricity, giving rise to a three equivalent formalisms \cite{Jimenez:2019woj}. 

When curvature and non-metricity vanish, the resulting theory is the so called \emph{Teleparallel Equivalent of General Relativity} (TEGR) (see \emph{e.g.} \cite{Maluf:2013gaa} for further details); when curvature and torsion vanish we have the so called \emph{Symmetric Teleparallel Equivalent of General Relativity} (STEGR) \cite{BeltranJimenez:2017tkd}. More precisely, in the former theory, it is possible to define two rank-three tensors of the form $T_{\,\,\,\, \mu \nu}^{\alpha} \equiv 2 \Gamma_{\,\,\,\, [\mu \nu]}^{\alpha}$ and $K^{\rho}_{\,\,\,\,\mu\nu}\equiv\frac{1}{2}g^{\rho\lambda}\bigl(T_{\mu\lambda\nu}+T_{\nu\lambda\mu}+T_{\lambda\mu\nu}\bigr)$, respectively called \emph{torsion tensor} and \emph{contorsion tensor}. Both of them  identically vanish in GR. In this way, by means of the \emph{superpotential} $S^{p \mu \nu} \equiv K^{\mu \nu p}-g^{p \nu} T_{\,\,\,\,\,\,\, \sigma}^{\sigma \mu}+g^{p \mu} T_{\,\,\,\,\,\,\, \sigma}^{\sigma \nu}$, one can define the torsion scalar as $T \equiv T^{p \mu \nu} S_{p \mu \nu}$, so that the TEGR action can be chosen to be linearly proportional to $T$. Interestingly, the action thus constructed, turns out to be equivalent to the Hilbert-Einstein one, up to a boundary term \cite{Aldrovandi:2013wha}. Moreover, TEGR can be recast as a gauge theory with respect to the translation group in the locally flat space-time \cite{Aldrovandi:2013wha}. 

Similar considerations apply for STEGR, where the definition of the \emph{non-metricity tensor} $Q_{\rho \mu \nu} \equiv \nabla_{\rho} g_{\mu \nu} \neq 0$ and the \emph{disformation tensor} $L^{\rho}_{\,\,\,\,\mu\nu}\equiv\frac{1}{2}g^{\rho\lambda}\bigl(-Q_{\mu \nu \lambda}-Q_{\nu \mu \lambda} + Q_{\lambda\mu\nu}\bigr)$, together with $Q_\mu \equiv Q_{\mu \,\,\,\, \lambda}^{\,\,\,\lambda}$ and $\tilde{Q}_{\mu} \equiv Q_{\alpha \mu}^{\, \, \, \, \, \, \alpha}$, allow to define the non-metricity scalar as: $Q \equiv - \frac{1}{4} Q_{\alpha \mu \nu} \left[- 2 L^{\alpha \mu \nu}+  g^{\mu \nu} \left(Q^\alpha - \tilde{Q}^\alpha \right) - \frac{1}{2} \left(g^{\alpha \mu} Q^\nu + g^{\alpha \nu} Q^\mu  \right)\right]$ \cite{BeltranJimenez:2017tkd}. Also here, the STEGR action, chosen to be linearly dependent on the non-metricity scalar, is equivalent to  GR and TEGR, up to a boundary term. Modifications of the STEGR action will not be considered in this work; for details on the fundamental structure of the theory and possible applications see \emph{e.g.}  Refs.~\cite{BeltranJimenez:2019tme, DAmbrosio:2021zpm}. 

Here we consider an extension of the TEGR action, containing a function of the torsion. This can allow to address the problems suffered by GR at large scales. Notice that, although GR and TEGR are dynamically equivalent, $f(R)$ gravity differs from $f(T)$ gravity. For instance, as the former leads to fourth-order field equations, the latter provides second-order equations with respect to the metric. This makes $f(T)$ gravity easy to handle  from a mathematical point of view.

In this paper, we want to discuss the problem of Minisuperspace Quantum Cosmology in metric and affine formulations of gravity theories. In particular, we will consider some metric and affine extensions. The aim is to show that, if the related cosmological models exhibit Noether symmetries, it is possible to interpret  solutions under the standard of the Hartle criterion and then achieve observable universes.
 
The paper is organized as follows:  Sec. \ref{QC1} is devoted to  the main features of ADM formalism, Quantum Cosmology and the relation between the latter and the Noether symmetries. In Sec. \ref{QCf(R)}, \ref{QCf(T)} and \ref{QCf(G)} we apply the Minisuperspace Quantum Cosmology formalism to $f(R)$, $f(T)$ and $f(\Gi)$ models, respectively. In Sec. \ref{concl},  we discuss the results and draw conclusions.  

\section{Quantum Cosmology and Noether Symmetries}
\label{QC1}
As mentioned in the Introduction, the ADM formalism can represent a useful tool in the context of Quantum Cosmology. Here we briefly summarize the foundations of the approach and provide a link between the latter and the Noether theorem. Generally, the most general form of the Hilbert-Einstein action includes the extrinsic scalar curvature $K$ \cite{DeWitt:1967yk,Thiemann:2007pyv}, defined as the contraction of the three-dimensional spatial curvature tensor $K_{ij}$ with the three-dimensional metric $h^{ij}$. Here middle indexes label the three-dimensional space. By defining a set of coordinates $X^{\alpha}$, the \emph{deformation vector} $N^\alpha = \dot{X}^\alpha$ can be decomposed in terms of the lapse function $N$ and the shift vector $N^i$ as:
\begin{equation}
N^\alpha = N n^\alpha + N^i X_i^\alpha\,,
\end{equation}
where $X_i^\alpha$ is a tangent vector basis characterizing each point of the three-dimensional surface and $n^\alpha$ a unitary vector satisfying the relations
\begin{equation}
g_{\mu \nu} X_i^\mu n^\nu = 0\,, \;\;\;\;\;\; g_{\mu \nu} n^\mu n^\nu = -1\,.
\label{vectors}
\end{equation}
Therefore, the metric can be recast as:
\begin{equation}
g_{\mu \nu} = \left(
\begin{matrix}

(N^2 - N_i N^i) & N_j 
\\
N_j & -h_{ij}

\end{matrix}
\right),
\end{equation}
by means of which the Lagrangian density reads:
\begin{equation}
\mathscr{L}=\frac{\kappa}{2} \sqrt{|h|} N\left(K^{i j} K_{i j}-K^{2}+{ }^{(3)} R\right) + t. d. \,,
\label{lagradens}
\end{equation}
with $h$ being the determinant of $h_{ij}$ and $^{(3)}R$ the intrinsic three-dimensional curvature. From Eq. \eqref{lagradens},  three conjugate momenta follow, respectively related to the shift vector, the lapse function and the three-dimensional metric:
\begin{eqnarray}
&& \pi \equiv \frac{\delta \mathscr{L}}{\delta \dot{N}}=0\,, \quad \pi^{i} \equiv \frac{\delta \mathscr{L}}{\delta \dot{N}_{i}}=0\,, \nonumber \\ 
&& \pi^{i j} \equiv \frac{\delta \mathscr{L}}{\delta \dot{h}_{i j}}=\frac{\kappa \sqrt{|h|}}{2}\left(K h^{i j}-K^{i j}\right).
\label{conjmom}
\end{eqnarray}
As a consequence, the Hamiltonian can be obtained by using Eq. \eqref{conjmom} and performing the Legendre transformation of the Lagrangian \eqref{lagradens}, so that we have:
\begin{equation}
\mathscr{H}=\pi^{i j} \dot{h}_{i j}-\mathscr{L}\,,
\end{equation}
constrained by the relations
\begin{equation}
\begin{cases}
\displaystyle \dot{\pi} = -\{\mathcal{H}, \pi \} = \frac{\delta \mathcal{H}}{\delta N} = 0\,,
\\
\displaystyle \dot{\pi}^i = -\{\mathcal{H}, \pi^i \} =  \frac{\delta \mathcal{H}}{\delta N_i} = 0\,.
\end{cases} 
\label{vectors2}
\end{equation}
In the above equations, it is $\mathcal{H} = \int \mathscr{H} \, d^3 x$\,. In the canonical quantization procedure, the momenta \eqref{conjmom} turn into the operators
\begin{equation}
\hat{\pi}=-i \frac{\delta}{\delta N}\,, \quad \hat{\pi}^{i}=-i \frac{\delta}{\delta N_{i}}\,, \quad \hat{\pi}^{i j}=-i \frac{\delta}{\delta h_{i j}}\,,
\label{primaryconstr}
\end{equation}
while the Poisson brackets \eqref{vectors2} turn into commutators. Finally, we obtain
\begin{equation}
\begin{cases}
[\hat{h}_{ij}(x), \hat{\pi}^{kl} (x')] = i \; \delta^{kl}_{ij} \; \delta^3 (x-x')\,,
\\
\delta^{kl}_{ij} = \frac{1}{2} (\delta_i^k \delta_j^l + \delta_i^l \delta_j^k)\,,
\\
[\hat{h}_{ij}, \hat{h}_{kl}] = 0\,,
\\
[\hat{\pi}^{ij}, \hat{\pi}^{kl} ] = 0\,.
\end{cases}
\end{equation}
Most importantly, from Eqs.~\eqref{vectors2}, it is 
\begin{equation}
\hat{{\cal{H}}}| \psi> = 0\,,
\label{WDWH}
\end{equation}
with $\psi$ being the wave function.  Eq. \eqref{WDWH} can be recast in terms of dynamical variables and momenta, leading to the WDW equation \cite{Montani}, which, in the case of  GR,  yields 
\begin{equation}
\left(\nabla^2 - \frac{\kappa^2}{4} \sqrt{|h|} \; ^{(3)} R \right)|\psi> = 0\,,
\label{constraints}
\end{equation}
with the operator $\nabla^2$ being defined as
\begin{equation}
\nabla^2 = \frac{1}{\sqrt{|h|}}\left(h_{i k} h_{j l}+h_{i l} h_{j k}-h_{i j} h_{k l}\right) \frac{\delta}{\delta h_{ij}}\frac{\delta}{\delta h_{kl}}\,.
\end{equation}
Eq. \eqref{constraints} shows that the ADM formalism relies on an infinite-dimensional superspace made of all possible 3-metrics, due to which any predictive power is inevitably lost. Moreover, as discussed above, the probabilistic interpretation of the wave function does not apply in this case, mainly because the scalar product $ \int \psi^* \psi \; dx^3$ is not positive-definite. As a consequence, an infinite-dimensional Hilbert space cannot be considered. In what follows we show that the ADM formalism can be suitably applied to cosmology, where the superspace can be reduced to a minisuperspace of configurations where the WDW equation, under some constraints, can be exactly solved. More precisely, the existence of Noether symmetries allows to introduce cyclic variables into the system, thanks to which the Hamiltonians assume  handy expressions and classical trajectories can be recovered.

Let us then introduce the main features of the Noether Theorem and how it can be  used as a method to select viable cosmological models which, eventually, result observable universes. To this purpose, we consider a transformation, involving coordinates and fields,  which is a symmetry for the Lagrangian density, namely
\begin{equation}
\begin{cases}
\tilde{x}^a = x^a - \xi^a 
\\
\tilde{\phi}^i = \phi^i - \eta^i \,,
\end{cases}
\label{trasformazioni coordinate}
\end{equation}
whose related first prolongation of Noether's vector can be written as:
\begin{equation}
 X^{[1]} = \xi^a \partial_a + \eta^i \frac{\partial }{\partial \phi^i} + (\partial_a \eta^i - \partial_a \phi^i \partial_b \xi^b) \frac{\partial}{\partial (\partial_a \phi^i)}.
\label{prolungamento generalizzato}
\end{equation}
The Noether Theorem states that, if the transformation \eqref{trasformazioni coordinate} is a symmetry for the Lagrangian, then there exists a gauge function $ g^a = g^a (x^a, \phi^i)$ satisfying the condition \cite{Dialektopoulos:2018qoe}
\begin{equation}
X^{[1]} \La + \partial_a \xi^a \La = \partial_a g^a
\label{theorem 2}
\end{equation}
and the quantity
\begin{equation}
j^a = \frac{\partial \La}{\partial (\partial_a \phi^i)} \eta^i - \frac{\partial \La}{\partial (\partial_a \phi^i)} \partial_b \phi^i \, \xi^b+ \La \xi^a - g^a,
\label{noethcurr1}
\end{equation}
is an integral of motion. For internal symmetries, Eqs. \eqref{prolungamento generalizzato} and \eqref{noethcurr1} reduce to:
\begin{eqnarray}
&&X = \eta^i \frac{\partial }{\partial \phi^i}  + \partial_a \eta^i \frac{\partial }{\partial (\partial_a \phi^i)} ,\label{j,x,internal} 
\\
&& j^a = \frac{\partial \La}{\partial (\partial_a \phi^i)} \eta^i.
\label{jinternal}
\end{eqnarray}
Consequently, setting $g^a = 0$, Eq. \eqref{theorem 2} becomes
\begin{equation}
X \La = 0
\end{equation}
and can be recast in terms of the Lie derivative along the flux of the vector $X$ as
\begin{equation}
L_X \La = 0.
\end{equation}
For further details about Noether's theorem and related applications see \emph{e.g.} \cite{Dialektopoulos:2018qoe, Acunzo:2021gqc, Bajardi:2020xfj}. It is worth pointing out that the conserved quantity \eqref{jinternal} can be used to properly change the minisuperspace variables if the related point-like Lagrangian is cyclic. Thanks to Noether's Theorem, it is possible to find a methodical procedure aimed at finding suitable new coordinates. To this purpose, let us assume that there exists a transformation which allows to introduce a variable $\psi^1$, whose related conjugate momentum is an integral of motion, that is
\begin{equation}
\frac{\partial \La}{\partial (\partial_a \psi^1)} = \pi^a_{\psi^1} = \text{const}.
\end{equation}
From Eq. \eqref{j,x,internal} we notice that $\psi^1$ is a constant of motion only if the related infinitesimal generator is equal to $1$. Indeed, considering the general change of variables $\phi^i \to \psi^i(\phi^j)$, such that $\psi^1$ is cyclic, the infinitesimal generators can be recast as:
\begin{equation}
\eta^i \frac{\partial}{\partial \phi^i}= \eta^i \frac{\partial \psi^j}{\partial \phi^i} \frac{\partial}{\partial \psi^j} = i_X d\psi^j \frac{\partial}{\partial \psi^j}.
\end{equation}
In this way, the Noether vector $X$ and the conserved quantity $j^a$, written in terms of the new variables, read
\begin{eqnarray}
X' = \eta^i \frac{\partial}{\partial \phi^i} + \partial_a \eta^i \frac{\partial}{\partial (\partial_a \phi^i)} = (i_X d\psi^k) \frac{\partial}{\partial \psi^k} + \partial_a (i_X d\psi^k)\frac{\partial}{\partial (\partial_a \psi^k)}, \nonumber
\label{generatore inner}
\end{eqnarray}
\begin{equation}
j^{a} = \eta^i \frac{\partial \La}{\partial (\partial_a \psi^i)} = i_X d\psi^i \frac{\partial \La}{\partial (\partial_a \psi^i)},
\label{nonext cons q}
\end{equation}
where $(i_X d\psi^k)$ is the inner derivative. Requiring the conserved quantity to be equal to the conjugate momentum of $\psi^1$, the following conditions must hold:
\begin{equation}
i_X d\psi^1 =\eta^j \frac{\partial \psi^1}{\partial \phi^j} = 1, \;\;\;\;\;\;\; i_X d \psi^i = \eta^j \frac{\partial \psi^i}{\partial \phi^j} = 0, \;\;\;\;\;\;\;\,\,\, i \neq 1,
\label{relzioni per cambio variabili}
\end{equation}
so that, from Eq. \eqref{nonext cons q}, one gets
\begin{equation}
j^a = \eta^i \frac{\partial \La}{\partial (\partial_a \phi^i)} = \frac{\partial \La}{\partial (\partial_a \psi^1)} \quad \to \quad \pi^a_{\psi^1} = \text{constant}.
\label{momento zeta}
\end{equation}
Therefore,  the conjugate momentum $\pi^a_{\psi^1}$ of the new cyclic variable $\psi^1$ turns exactly into the Noether current, as expected by construction.

Notice that the condition $X' \La' = X \La = 0$ holds independently of the variables considered, so that any Noether symmetry is  preserved under the change of variables. Nonetheless, it is worth remarking that the change of variables in Eq. \eqref{relzioni per cambio variabili} is not unique, but infinite possible field transformations can occur. Therefore, in order to reduce  dynamics, new variables must be chosen carefully. 

If the variables in the minisuperspace depend only on the general parameter $t$, like in cosmology, Eqs. \eqref{prolungamento generalizzato}, \eqref{noethcurr1}, \eqref{j,x,internal} and \eqref{jinternal} provide:
\begin{eqnarray}
&& X^{[1]} = \dot{\xi} \partial_t + \eta^i \frac{\partial }{\partial q^i} + (\dot{\eta^i} - \dot{q}^i \dot{\xi}) \frac{\partial}{\partial \dot{q}^i}, 
\\
&& X = \eta^i \frac{\partial }{\partial q^i}  + \dot{\eta}^i \frac{\partial }{\partial \dot{q}^i} , 
\\
&& j = \frac{\partial \Lagr}{\partial \dot{q}^i} \eta^i - \frac{\partial \Lagr}{\partial \dot{q}^i} \dot{q}^i \, \xi + \Lagr \xi - g, 
\\
&& j = \frac{\partial \Lagr}{\partial \dot{q}^i} \eta^i,
\end{eqnarray}
with $q^i$ being the variables in the cosmological minisuperspace and $\Lagr$ the Lagrangian. Therefore, the change of variables $q^i \to Q^i(q^j)$, which allows to introduce a constant of motion, can be recast as:
\begin{equation}
i_X dQ^1 =\eta^j \frac{\partial Q^1}{\partial q^j} = 1, \;\;\;\;\;\;\; i_X d Q^i = \eta^j \frac{\partial Q^i}{\partial q^j} = 0, \;\;\;\;\;\;\;\,\,\, i \neq 1.
\label{relzioni per cambio variabili1}
\end{equation}
After summarizing the main features of ADM formalism and Noether Theorem, it is worth investigating the deep connection between them, introducing the so called \emph{Hartle criterion}. To this purpose, let us notice that Eq. \eqref{relzioni per cambio variabili1} permits to recast the conjugate momenta of the cyclic variables as
\begin{equation}
j_i \equiv \pi^i_{Q^i} = \frac{\partial \Lagr}{\partial \dot{Q}^i}.
\label{jcosmo}
\end{equation}
As a consequence, if the cosmological point-like Lagrangian enjoys $m$ symmetries with related conserved quantities $j_1, j_2 \dots j_m$, Eq. \eqref{WDWH} together with  \eqref{primaryconstr} yield the following system of $m+1$ differential equations:
\begin{equation}
\begin{cases}
\hat{\mathscr{H}} |\psi>=0
\\
- i \partial_1 |\psi> = j_1 |\psi>
\\
- i \partial_2 |\psi> = j_2 |\psi>
\\
\mathrel{\vdots}
\\
- i \partial_m |\psi> = j_m |\psi>.
\end{cases}
\label{systemconstraint1}
\end{equation}
Notice that, thanks to the change of variables \eqref{relzioni per cambio variabili1}, suggested by the Noether symmetries, the above system can be suitably integrated. Indeed, the integrals of motion $j_i$ allow to reduce dynamics and generally provide a wave function of the form \cite{Capozziello:1999xr}:
\begin{equation}
|\psi> = e^{i j_k Q^k} |\chi(Q^\ell)>, \qquad m < \ell < n,
\end{equation}
where $m$ are the variables with symmetries (integrals of motion),  $\ell$ are the variables  with no symmetries and $n$ the minisuperspace dimension. Interestingly, in the semiclassical limit, the existence of symmetries leads to oscillatory solutions of the WDW equation. The latter, according to the Hartle criterion, are related to observable universes. Specifically, Hartle proposal relies on the analogy with non-relativistic Quantum Mechanics, where the oscillating wave function generally describes a classically permitted region, unlike the exponential wave function which labels the quantum region. Similar arguments can be also applied within the context of Quantum Cosmology, where the existence of Noether symmetries  assures that the Hartle criterion  can be  related to classical trajectories. In other words, the oscillations of some components of the wave function mean correlations among variables while the exponential behavior means no correlation. Moreover, in the \emph{Wentzel-Kramers-Brillouin} (WKB) limit, the wave function can be linked to the gravitational action $S$ by means of the relation $\psi(h_{ij}, \phi) \sim e^{i S}$. Therefore, using the Hamilton-Jacobi equations $\displaystyle \frac{\partial S}{\partial q^a} = \pi_a$, it is possible to get the dynamics for the generic variable $q^a$.

In other words, the importance of Noether symmetries in Quantum Cosmology is twofold: on the one hand, symmetries allow to reduce the dynamical system of differential equations arising from the ADM formalism, to find analytic solutions and to link such solutions to observable universes. On the other hand, the wave function can be recast in terms of the cosmological action to recover the Euler-Lagrange equations with respect to the new variables. This permits to suitably find exact solutions.  The physical interpretation of such solutions is related to the Hartle criterion. For further readings on Quantum Cosmology and applications to  theories of gravity see \emph{e.g.} \cite{Faraoni:2010pgm, Unruh:1989db, Schwinger:1963re}.

The results discussed above make Quantum Cosmology an important connection  between classical and quantum aspects of gravity; while waiting for a complete and self-consistent theory of Quantum Gravity,  applications to cosmology give interpretative approach capable of reducing the infinite-dimensional superspace coming from the ADM formalism to minisuperspaces where the equations of motion can be integrated and, eventually, interpreted. 

\section{Minisuperspace Quantum Cosmology in $f(R)$ gravity}
\label{QCf(R)}
A first application of the above considerations can be developed for  $f(R)$ gravity described by the action
\begin{equation}
S = \int \sqrt{-g} f(R) \, d^4x.
\label{f(R) action}
\end{equation}
We will select the functional form of the above action by the Noether Symmetry Approach and will get the explicit expression of the point-like Lagrangian, which can be rendered cyclic by applying conditions in Eq. \eqref{relzioni per cambio variabili1}. Finally, we will find the Hamiltonian in terms of the new variables and will solve the cosmological WDW equation for this modified model. This procedure leads to the Wave Function of the Universe, by means of which classical trajectories can be suitably obtained. We will also show that the Hartle criterion is recovered, in agreement with the above discussion.

By varying Eq. \eqref{f(R) action} with respect to the metric, one gets:
\begin{equation}
G_{\mu \nu} = \frac{1}{f_R(R)} \left\{\frac{1}{2} g_{\mu \nu} \left[f(R) - R f_R(R) \right] + f_R(R)_{; \mu ; \nu} - g_{\mu \nu} \Box f_R(R) \right\},
\label{field equation f(R)}
\end{equation}
being $G_{\mu \nu}$ the Einstein tensor $\displaystyle G_{\mu \nu} = R_{\mu \nu} - \frac{1}{2} g_{\mu \nu} R$ and $f_R(R) $ the first derivative of $f(R)$ with respect to $R$. Notice that when $f_R = 1$, Einstein field equations are recovered. 
Also, the RHS can be understood as an effective energy-momentum tensor provided by the geometry, capable of reproducing the  Dark Energy behavior without any further material ingredient \cite{Capozziello:2002}.
 
Before applying the Noether Symmetry Approach, let us  find the related point-like Lagrangian in a cosmological spatially-flat background of the form $ds^2 = dt^2 - a(t) d$\textbf{x}$^2$, where $a(t)$ is the scale factor. Using a Lagrange multiplier $\lambda$, and considering the cosmological expression of the scalar curvature, the action can be written as
\begin{equation}
S = \int \left[a^3 f(R) - \lambda \left(R + 6 \frac{\ddot{a}}{a}+ 6 \frac{\dot{a}^2}{a^2} \right) \right] dt,
\label{f(R)actionionit}
\end{equation}
where we integrated the three-dimensional hypersurface. By varying the action with respect to the  curvature scalar,  it is possible to find the Lagrange multiplier $\lambda$, that is:
\begin{equation}
\frac{\delta S}{\delta R} = a^3 f_R(R) - \lambda = 0, \,\,\,\,\,\,\,\,\,\,\,\, \lambda = a^3 f_R(R).
\label{lagrangemult}
\end{equation} 
Replacing the result in Eq. \eqref{lagrangemult} into the action \eqref{f(R)actionionit} and integrating out second derivatives, the canonical point-like Lagrangian turns out to be \cite{Capozziello:2008ch}
\begin{equation}
\Lagr (a,\dot{a}, R,\dot{R})= a^3 \left[ f(R) - R f_R(R) \right] + 6 a \dot{a}^2 f_R(R) + 6 a^2 \dot{a} \dot{R} f_{RR}(R).
\label{Lagr f(R)}
\end{equation}
The corresponding Euler-Lagrange equations are
\begin{equation}
\begin{cases}
\displaystyle 6 a^2 \dot{a} \dot{R} f_{RR}(R) + 6 a \dot{a}^2 f_R(R) - a^3[f(R) - R f_R(R)] = 0
\\
\displaystyle \dot{R}^2 f_{RRR}(R)  + \ddot{R} f_{RR}(R) + \frac{\dot{a}^2}{a^2} f_R(R) + 2 \frac{\ddot{a}}{a} f_R(R) 
- \frac{1}{2} [f(R) - R f_R(R)] + 2\frac{\dot{a}}{a} \dot{R} f_{RR}(R) = 0
\\
\displaystyle R = - 6 \left( \frac{\ddot{a}}{a}+  \frac{\dot{a}^2}{a^2} \right).
\end{cases}
\label{EL f(R)}
\end{equation}
The first equation is the energy condition $E_\Lagr = 0$, corresponding to the (0,0) component of the field equations, namely the modified first Friedmann equation. The second is the equation with respect to the scale factor; the third is the Euler-Lagrange equation with respect to the scalar curvature, which provides again the cosmological expression of $R$ derived from the Lagrange multiplier.
In the considered minisuperspace, the Noether vector assumes the form 
\begin{eqnarray}
&& X^{[1]} = \xi(a,R,t) \partial_t + \alpha(a,R,t) \partial_a + \beta(a,R,t) \partial_R  
\\
&& +  \left(\dot{\alpha}(a,R,t) - \dot{\xi}(a,R,t) \dot{a} \right) \partial_{\dot{a}} + \left(\dot{\beta}(a,R,t) - \dot{\xi}(a,R,t) \dot{R} \right) \partial_{\dot{R}},\nonumber
\end{eqnarray}
so that the application of the Noether identity \eqref{theorem 2}, yields the following possible solution
\begin{equation}
\begin{cases}
\displaystyle  {\cal{X}}= \frac{\alpha_0}{a}  \partial_a -  2 \alpha_0 \frac{R}{a^2}  \partial_R 
\\
\displaystyle j = 9 \alpha_0 f_0 (2 R \dot{a} +  a \dot{R}) R^{- \frac{1}{2}}, \,\,\,\,\,\, f(R) = f_0 R^{\frac{3}{2}},
\end{cases}
\label{solution noether f(R)}
\end{equation}
where $\mathcal{X}$ is the symmetry generator. For the entire set of solutions see \cite{Capozziello:2008ch, Benetti}. Notice that Eqs. \eqref{solution noether f(R)} describes an internal symmetry, thus the infinitesimal generator $\xi$ vanishes identically. This means that Eq. \eqref{solution noether f(R)} is also a solution of the vanishing Lie derivative condition and, consequently, the  change of variables in Eq. \eqref{relzioni per cambio variabili1} can be adopted.
\subsection{The WDW equation and the  Wave Function of the Universe}
A suitable Minisuperspace Quantum Cosmology can be constructed for  the function $\displaystyle f(R) = f_0 R^{\frac{3}{2}}$, selected by  the Noether symmetries.  Before considering the ADM formalism and finding the related Hamiltonian, we use the  Noether Approach to reduce  dynamics, by introducing a cyclic variable in the minisuperspace. To this purpose, Eq. \eqref{relzioni per cambio variabili} permits to pass from the minisuperspace $\mathcal{S}\{a,R\}$ to $\mathcal{S}'\{z,w\}$, by means of the following system of differential equations:
\begin{equation}
\begin{cases}
\alpha \partial_a z(a,R) + \beta \partial_R z(a,R) = 1
\\
\alpha \partial_a w(a,R) + \beta \partial_R w(a,R) = 0,
\end{cases}
\label{systemchangevar}
\end{equation} 
with $z$ being the  cyclic variable. A possible  solution is
\begin{equation}
\begin{cases}
\displaystyle w = w_0 (a \sqrt{R})^\ell, \,\,\,\,\,\, z = \frac{a^2}{2 \alpha_0}
\\
\displaystyle  a = \sqrt{2 \alpha_0 z}, \,\,\,\,\,\,\,\,\,\,\,\,\,\,\,\, R =\frac{1}{2 \alpha_0} z \, \left(\frac{w}{w_0} \right)^{\frac{2}{\ell}},
\end{cases}
\label{z,w f(R)}
\end{equation}
being $\ell$ and $w_0$ integration constants. Replacing Eq. \eqref{z,w f(R)} into Eq. \eqref{Lagr f(R)}, the Lagrangian turns out to be
\begin{equation}
\Lagr = \frac{f_0}{\ell w} \left(\frac{w}{w_0} \right)^{\frac{1}{\ell}} \left[18 \alpha_0 \dot{w} \dot{z} - w \ell \left(\frac{w}{w_0} \right)^{\frac{2}{\ell}} \right].
\end{equation}
Notice that the new equations of motion are simpler than   Eqs. \eqref{EL f(R)}, written in terms of the old variables. They read:
\begin{equation}
\begin{cases}
\displaystyle (\ell-1) \dot{w}^2 - \ell w \ddot{w} = 0
\\
\displaystyle \left(\frac{w}{w_0} \right)^{\frac{2}{\ell}}+ 6 \alpha_0 \ddot{z} = 0\,,
\end{cases}
\label{transformed EL f(R)}
\end{equation}
where clearly $z$ is the cyclic variable because there is no potential term depending on it.
The Legendre transformation $\mathcal{H} = \pi_i \dot{q}^i - \Lagr$ provides the Hamiltonian
\begin{equation}
\mathcal{H} = \frac{\ell }{18 \alpha_0 f_0 }  w \left(\frac{w}{w_0} \right)^{-\frac{1}{\ell}} \pi_w \pi_z  + 
 f_0 \left(\frac{w}{w_0} \right)^{\frac{3}{\ell}}.
\end{equation}
By promoting the momenta to operators, \emph{i.e.} $\pi_i \to -i \partial_i$, the primary and secondary constraints in Eqs. \eqref{WDWH} and \eqref{primaryconstr} read as:
\begin{equation}
\begin{cases}
\displaystyle \hat{\mathcal{H}} \psi =  \left[\frac{\ell }{18 \alpha_0 f_0 }  w \left(\frac{w}{w_0} \right)^{-\frac{1}{\ell}} \partial_w \partial_z  + 
 f_0 \left(\frac{w}{w_0} \right)^{\frac{3}{\ell}}\right] \psi = 0
\\
\\
\displaystyle \hat{\pi}_z \psi = - i \partial_z \psi = j_0 \psi,
\end{cases}
\end{equation}
where the former is the WDW equation. The solution of the above system yields the following wave function \cite{Capozziello:2012hm}
\begin{equation}
\psi(z,w) = \psi_0 \exp \left\{i \left[j_0 z - \frac{9 \alpha_0 f_0^2}{2 j_0} \left(\frac{w}{w_0} \right)^{\frac{4}{\ell}} \right]\right\}.
\label{WFf(R)}
\end{equation}
In the semiclassical limit, the wave function can be recast in terms of the action $S$ as 
\begin{equation}
\psi \sim e^{i S},
\label{WKBLIMIT}
\end{equation}
so that comparing Eq. \eqref{WKBLIMIT} with Eq. \eqref{WFf(R)}, the action turns out to be
\begin{equation}
S = j_0 z - \frac{9 \alpha_0 f_0^2}{2 j_0} \left(\frac{w}{w_0} \right)^{\frac{4}{\ell}}.
\end{equation}
It can be easily proven that the Hamilton-Jacobi equations $\displaystyle \frac{\partial S}{\partial q^i} = \pi_i$ exactly provide the same system as Eq. \eqref{transformed EL f(R)}. The related solution, after coming back to the old variables, reads \cite{Capozziello:2012hm}
\begin{equation}
a(t)=a_0 \left[c_{4} t^{4}+c_{3} t^{3}+c_{2} t^{2}+c_{1} t+c_{0}\right]^{1 / 2},
\label{cosmosol f(R)}
\end{equation}
with $c_i$ integration constants. It is worth noticing that, due to the oscillatory behaviour of the wave function \eqref{WFf(R)}, the Hartle criterion is recovered by the existence of the Noether symmetry. The cosmological solution, emerging after this process, is an \textit{observable universe} \eqref{cosmosol f(R)}. Other solutions of this type can be found, see 
\emph{e.g.},  \cite{Jamil:2011pv}.

\section{Minisuperspace Quantum Cosmology in $f(T)$ gravity}
\label{QCf(T)}
Let us  now consider  an action containing a function of the torsion scalar $T$, namely
\begin{equation}
S = \int e \, f(T) \, d^4x,
\label{f(T)action}
\end{equation}
defined as an extension  of TEGR. 
As discussed  in Sec.\ref{introd}, $T$ is the contraction of the superpotential with the torsion tensor. We assume the torsion scalar to be written in terms of the Weitzenb\"ock connection $\Gamma^\alpha_{\,\,\, \mu \nu} = e^\rho_a \partial_\mu e^a_\nu$, with $e^a_\mu$ being the tetrad fields. By this choice, the spin connection vanishes identically, but the Lorentz Invariance is formally preserved and the dynamics results unchanged \cite{Krssak:2015oua}.

By varying the action \eqref{f(T)action} with respect to the tetrad fields, one gets the following field equations:
\begin{eqnarray}
\frac{1}{e} \partial_\mu(h \; e^\rho_a S\,_\rho^{\;\; \mu \nu}) f_T(T) - e^\lambda_a T\,^\rho_{\;\; \mu \lambda} S\,_\rho^{\;\; \nu \mu} f_T(T) + e^\rho_a S_\rho^{\;\; \mu \nu}(\partial_\mu T) f_{TT}(T) + \frac{1}{4} e^\nu_a f(T) = 0,
\label{equazioni di campo f(T)}
\end{eqnarray}
with $f_T(T)$ being the first derivative of $f(T)$ with respect to $T$ and $e$ the tetrad fields determinant. Unlike the $f(R)$ model, assigning a cosmological spatially-flat space-time is not sufficient to uniquely determine the form of the tetrad fields. To overcome this issue, some criterions of "good" and "bad" tetrads can be used, as shown for example in \cite{Tamanini:2012hg}. In the applications to cosmology, we can use the simplest choice among all possible tetrads leading to a spatially-flat cosmological universe, namely $e^a_\mu = diag(1,a(t), a(t), a(t))$. Consequently, the field equations \eqref{equazioni di campo f(T)} take the form 
\begin{eqnarray}
&& T f_T(T) - \frac{1}{2} f(T) = 0\,,   \label{f(T)cosmo1}
\\
&& 2T f_{TT}(T)   + f_T(T) = 0\,.
\label{f(T)cosmo2}
\end{eqnarray}
The same system of differential equations arises from the Euler-Lagrange equations coming from the point-like Lagrangian, which, in turn, can be obtained by means of the Lagrange Multipliers Method. More precisely, using the cosmological expression of the torsion scalar 
\begin{equation}
T = -6\frac{\dot{a}^2}{a^2},
\label{vincolo T}
\end{equation}
the action \eqref{f(T)action} can be recast as:
\begin{equation}
S = \int \left[a^3 f(T) - \lambda \left(T + 6 \frac{\dot{a}^2}{a^2} \right) \right] dt.
\end{equation}
Also here, the Lagrange multiplier $\lambda$ is straightforwardly provided by the variation of the action with respect to the torsion scalar, that is
\begin{equation}
\frac{\delta S}{\delta T} = 0 \,\,\, \to \,\,\, \lambda = a^3 f_T(T),
\end{equation}
so that we have
\begin{equation}
\mathcal{L}(a,\dot{a}, T)= a^3 [f(T) - T f_T (T)] - 6 a \dot{a}^2 f_T(T).
\label{Lagra f(T)}
\end{equation}
As mentioned above, the dynamical system of Euler-Lagrange equations turns out to be exactly equivalent to that provided by the field equations \eqref{f(T)cosmo1} and \eqref{f(T)cosmo2}. This system, however, can be analytically solved after selecting the form of the function $f(T)$. To this purpose, as in the previous section, we consider the Noether Symmetry Approach and develop the corresponding quantum cosmological model.

Let us  start from Lagrangian \eqref{Lagra f(T)} and select the unknown function by Noether's approach. In the  considered two-dimensional minisuperspace, that is $\mathcal{S} = \{a,T\}$, the first prolongation of the Noether vector takes the form:
\begin{equation}
X^{[1]}= \xi \partial_t + \alpha \partial_a + \beta \partial_T + \left( \dot{\alpha} - \dot{\xi} \dot{a} \right) \partial_{\dot{a}},
\end{equation}
where $\alpha = \alpha(a,T)$ and $\beta = \beta(a,T)$ are the components of the above  $\eta^i$ defined in \eqref{prolungamento generalizzato}. By applying the Noether symmetry existence condition \eqref{theorem 2} and equating to zero  terms with the same time derivatives, we get a system of partial differential equations. A possible solution is
\begin{equation}
\begin{cases}
\displaystyle {\cal{X}} =\left[\frac{\xi_0}{2 k-1} \right]t \partial_{t}+\left[\frac{\xi_0}{3}\, a+\alpha_{0} a^{1-\frac{3}{2 k}}\right] \partial_{a} 
+ \left[\frac{1}{k}\left(\beta_1 - 3 \frac{\alpha_0} a^{-\frac{3}{2 k}}\right)+\frac{\xi_0}{2 k-1}+\beta_2\right] T \partial_{T}, \,\,\,\,\,\,\,\, k \neq \frac{3}{2}, \frac{1}{2}
\\
\displaystyle j_0 = -12 f_{0} k\left(\xi_0 a^{2}+ \alpha_0 a^{2-\frac{3}{2 k}}\right) T^{k-1} \dot{a} .
\end{cases}
\label{generatorf(T)}
\end{equation}
 See \cite{Bajardi:2021tul} for other solutions. To  introduce a cyclic coordinate by  the procedure  in Eqs. \eqref{relzioni per cambio variabili1}, we set $\xi_0 = \beta_1= \beta_2 = 0$, so that internal symmetries are selected. In this way, the components of the function $\eta^i$ read:
\begin{equation}
\alpha = \alpha_0 a^{1 - \frac{3}{2k}}, \;\;\;\;\;\;\;\;\;\;\;\;\;\;\; \beta = \frac{-3\alpha_0}{k}T a^{- \frac{3}{2k}}. \label{alpha e beta}
\end{equation}
By means of Eqs. \eqref{relzioni per cambio variabili1}, it is possible to introduce a cyclic variable in the minisperspace ${\cal{S}}= \{ a,T\}$. Specifically, Eqs. \eqref{relzioni per cambio variabili1} provide:
\begin{equation}
\begin{cases}
\displaystyle \alpha \partial_a z + \beta \partial_T z = 1
\\
 \displaystyle \alpha \partial_a w + \beta \partial_T w = 0,
\end{cases}
\label{cambio variabili f(T)}
\end{equation}
which, using the solution \eqref{alpha e beta}, reduces to 
\begin{equation}
\begin{cases}
 \displaystyle z = \frac{2k}{3\alpha_0}a^{\frac{3}{2k}}, \;\,\,\,\, \to \,\,\,\, a = \left(\frac{3\alpha_0 z}{2k} \right)^{\frac{2k}{3}}
\\
\displaystyle w = a^3 T^k, \,\,\,\,\,\,\,\,\,\,\, \to \,\,\,\, T = w^{\frac{1}{k}} \left(\frac{3\alpha_0 z}{2k} \right)^{-2},
\end{cases}
\end{equation}
so  the  Lagrangian with cyclic variable reads
\begin{equation}
{\cal L} = w(1-k) -6k \alpha_0^2 \dot{z}^2 w^{\frac{k-1}{k}} \;,
\label{new lagr f(T)}
\end{equation}
and $z$ is cyclic as expected. The set of Euler-Lagrange equations coming from the above Lagrangian is 
\begin{eqnarray}
&&  k w \ddot{z} + \dot{w} \dot{z} (k-1) = 0, \label{ELz}
\\
&& w = - 6^k  \alpha_0^{2k} \dot{z}^{2k}. \label{ELw}
\end{eqnarray} 
Following the same steps as $f(R)$ gravity, we perform a Legendre transformation for  Lagrangian \eqref{new lagr f(T)}, that is: 
\begin{equation}
\displaystyle {\cal H} = w(k-1) + \frac{3 \pi^2_z}{24 k \alpha^2_0 w^{\frac{k-1}{k}}} \;.
\label{Hamf(T)}
\end{equation}
In the canonical quantization scheme, where the conjugate momenta and the Hamiltonian are recast in terms of differential operators, the primary constraints \eqref{primaryconstr} and \eqref{WDWH} yield the system:
\begin{equation}
\begin{cases}
\displaystyle \left[w(k-1) - \frac{3 \partial^2_z}{24 k \alpha^2_0 w^{\frac{k-1}{k}}}\right]\psi = 0
\\
\displaystyle i\partial_z \psi = j_0 \psi  \;,
\end{cases}
\end{equation}
where the first equation is the WDW equation  and the second is the momentum conservation equation. A  solution is:
\begin{equation}
\psi \sim \exp \left\{i\left[2 \alpha_0 w^{\frac{2k -1}{2k}} \sqrt{2k(k-1)}\right]z\right\} \;.
\label{psif(T)}
\end{equation}
Notice that, also here, the wave function is oscillating, confirming that the Hartle criterion holds for this model and observable universes are naturally provided. Moreover, in the WKB approximation, the wave function can be related to the gravitational action, so that the latter can be explicitly recast as:
\begin{equation}
\nonumber S = \left[2 \alpha_0 w^{\frac{2k -1}{2k}} \sqrt{2k(k-1)}\right]z.
\end{equation}
Using the Hamilton-Jacobi equations
\begin{equation}
\begin{cases}
\displaystyle \frac{\partial S}{\partial z} = \pi_z = j_0
\\
\displaystyle \frac{\partial S}{\partial w} = \pi_w = 0 \;,
\end{cases}
\label{ELWF}
\end{equation} 
we obtain the following analytic solution for $z$ and $w$
\begin{equation}
z(t) = z_0 t, \quad w =  - 6^k  \alpha_0^{2k} z_0^{2k},
\end{equation}
which, in terms of the old variables, becomes
\begin{equation}
\label{classical}
a(t) = a_0 t^{\frac{2k}{3}}, \; \;\;\; \; T(t) = -\frac{8 k^2}{3} \frac{1}{t^2} .
\end{equation}
The free parameter $k$ can be constrained by the energy condition, which is the first Friedmann equation. It is possible to show that the only admissible solution occurs for $k=1/2$. This means that the only cosmological solution which is compatible with Noether symmetries describes a stiff matter dominated epoch.

\section{Minisuperspace Quantum Cosmology in $f(\Gi)$ gravity}
\label{QCf(G)} 
As mentioned in the Introduction, Gauss--Bonnet cosmology has been recently taken into account because it  allows to reduce the complexity of the field equations and, at the same time, to solve some high-energy issues exhibited by GR. Moreover, once considering an action proportional to $R + f(\Gi)$, the function can be understood as an effective cosmological constant, with negligible contributions at the level of  Solar System. Therefore, in the limit $f(\Gi) \to 0$, GR is safely recovered. Nevertheless, it s possible to show that Einstein's  theory can be obtained even without imposing the GR limit as a requirement, namely when the action is only proportional to the function $f(\Gi)$. Specifically, in cosmological contexts, the model $f(\Gi) \sim \sqrt{\Gi}$ turns out to be dynamically equivalent to the scalar curvature \cite{Bajardi:2020osh}. Similar considerations also apply in a spherically symmetric background \cite{Bajardi:2019zzs}. Let us  start with the action
\begin{equation}
S = \int \sqrt{-g} f(\Gi) \, d^4x,
\label{ACTGB}
\end{equation}
whose variation with respect to the metric tensor provides \cite{Nojiri:2005jg, Bamba:2017cjr}:
\begin{eqnarray}
&&\frac{1}{2} g_{\mu \nu} f(\Gi) - \left(2R R_{\mu \nu} - 4 R_{\mu p} R^p_{\,\,\,\nu} + 2 R_\mu^{\,\,\, p \sigma \tau} R_{\nu p \sigma \tau} - 4 R^{\alpha \beta} R_{\mu \alpha \nu \beta}\right) f_\Gi(\Gi) +\nonumber
\\
&& + \left(2R \nabla_\mu \nabla_\nu +4 G_{\mu \nu} \Box - 4 R_{\{ \nu}^p \nabla_{\mu \}}\nabla_p + 4 g_{\mu \nu} R^{p \sigma} \nabla_p \nabla_\sigma - 4 R_{\mu \alpha \nu \beta} \nabla^\alpha \nabla^\beta \right) f_\Gi(\Gi) = 0 \;.
\label{FEGB}
\end{eqnarray}
In a cosmological spatially flat space-time, the Gauss--Bonnet invariant takes the form
\begin{equation}
\Gi = 24 \frac{\dot{a}^2 \ddot{a}}{a^3}.
\label{GBexprcosmo}
\end{equation}
It is easy to see that, when multiplied by $\sqrt{-g}$, Eq. \eqref{GBexprcosmo} turns out to be a boundary term. In order to apply the Noether Symmetry Approach and find the cosmological Hamiltonian, we adopt a Lagrangian description as in the previous sections. Therefore, we use the constraint \eqref{GBexprcosmo} to recast the action \eqref{ACTGB} in terms of Lagrange multipliers as
\begin{equation}
S = \int \left[a^3 f(\mathcal{G}) - \lambda \left\{ \mathcal{G} - 24 \frac{\dot{a}^2 \ddot{a}}{a^3} \right\}  \right] d^{4}x \;.
\label{azione con lambda}
\end{equation}
Using the variational principle and integrating out higher derivatives, the canonical point-like Lagrangian turns out to be:
\begin{equation}
\Lagr(a,\dot{a}, \Gi, \dot{\Gi}) =  a^3 [f(\Gi) - \Gi f_{\Gi}(\Gi)] - 8 \dot{a}^3 \dot{\Gi} f_{\Gi \Gi}(\Gi) .
\label{Lagr4dfG}
\end{equation}
The field equations \eqref{FEGB} are thus equivalent to the system of equations of motion coming from Eq. \eqref{Lagr4dfG}, which reads
\begin{equation}
\begin{cases}
a^3 [f(\Gi) - \Gi f_{\Gi}(\Gi)] + 24  \dot{a}^3 \dot{\Gi} f_{\Gi \Gi}(\Gi) = 0
\\
 \, a^2 [f(\Gi) - \Gi f_{\Gi}(\Gi)] + 8  \dot{a} [2 \dot{\Gi} \ddot{a} f_{\Gi \Gi}(\Gi) + \dot{a} \ddot{\Gi} f_{\Gi \Gi}(\Gi) + \dot{a} \dot{\Gi}^2 f_{\Gi \Gi \Gi}(\Gi)]
\\
 \displaystyle \Gi =24 \frac{\dot{a}^2 \ddot{a}}{a^3}.

\label{EL3}
\end{cases}
\end{equation}
The first equation is the energy condition. The second is the equation for the scale factor evolution and the third is the equation for the Gauss-Bonnet term coinciding with the Lagrange multiplier  \eqref{GBexprcosmo}.

In the two-dimensional minisuperspace $\mathcal{S} = \{a, \Gi\}$, the generator of transformations can be
\begin{equation}
\mathcal{X} = \xi(a,\Gi,t) \partial_t + \alpha(a,\Gi,t) \partial_a + \beta(a,\Gi,t) \partial_{\Gi}\,.
\end{equation} 
The transformation  is a symmetry,  if the condition \eqref{theorem 2} holds. Pursuing the same procedure as in Secs. \ref{QCf(R)} and \ref{QCf(T)}, the application of the Noether identity to Lagrangian \eqref{Lagr4dfG} provides a system of two differential equations:
\begin{equation}
\begin{cases}
\displaystyle 3 \al a^2 [ f(\Gi) - \Gi f'(\Gi)] - \be a^3 \Gi f''(\Gi) + \partial_t \xi a^3 [ f(\Gi) - \Gi f'(\Gi)] = 0
\\
\displaystyle 3 \partial_a \alpha 	 f''(\Gi)  + \be f'''(\Gi) - 3 \partial_t \xi\; f''(\Gi) + \partial_{\Gi} \be \; f''(\Gi) = 0
\\
\displaystyle \xi = \xi(t)\,, \quad \alpha = \alpha(a)\,, \quad \beta = \beta(\Gi) \quad g = g_0,
\end{cases}
\end{equation}
whose solution is \cite{Bajardi:2020osh}
\begin{equation}
\alpha = \alpha_0 a\,, \quad \beta = - 4 \xi_0 \Gi\,, \quad \xi = \xi_0 t + \xi_1\,, \,\, f(\Gi) = f_0 \Gi^{k}, \quad j_0 = \frac{\dot{a}^3}{\Gi^{3k}} ,
\label{QconsG}
\end{equation}
with the definitions
\begin{equation}
k \equiv \frac{3\alpha_0 + \xi_0}{4 \xi_0}, \qquad f_0 \equiv \frac{ 4 f_0 \xi_0}{3\alpha_0  + \xi_0}.
\label{kf0}
\end{equation}
Replacing the selected function into the Lagrangian \eqref{Lagr4dfG}, the latter takes the form:
\begin{equation}
\Lagr = -\frac{1}{3} \Gi^{k-2} \left[3 (k-1) a^3 \Gi^2 + 24 k (k-1) \dot{a}^3\dot{\Gi} \right].
\label{LagraGB1}
\end{equation}
Starting from this Lagrangian, it is possible to find the related Hamiltonian and the Wave Function of the Universe. 
\subsection{The WDW equation and Wave Function of the Universe}

As above, applying Eqs. \eqref{relzioni per cambio variabili1},  the minisuperspace can be reduced. Here,  setting $\xi = 0$, leads to trivial solutions. Therefore, only symmetries involving space-time translations can be selected  and this fact does not allow to use the procedure,  based on Eq. \eqref{relzioni per cambio variabili1}, as in previous sections. Nevertheless, starting from Lagrangian \eqref{LagraGB1}, it is still possible to write the Hamiltonian as:
\begin{equation}
\mathcal{H} = \frac{f_0}{k} \Gi^k a^3 + \pi_a \left(-\frac{\pi_\Gi}{8f_0} \Gi^{2-k} \right)^{\frac{1}{3}} \;,
\label{Hamiltoniana finale}
\end{equation}
with $\displaystyle \pi_a=\frac{\partial {\cal L}}{\partial \dot {a}}$ and $\displaystyle \pi_{\Gi}=\frac{\partial {\cal L}}{\partial \dot {\Gi}}$. In this form, the system cannot be immediately quantized due to the presence of the fractional exponent. However, we can still use the conserved quantity \eqref{QconsG} to reduce the minisuperspace and to get a suitable Hamiltonian. More precisely,
the momentum $\pi_{\Gi}$ can be rewritten in terms of $j_0$ as:
\begin{equation}
\pi_{\Gi} = -8 f_0 j_0 \Gi^{4k-2} \;,
\end{equation}
so that the Hamiltonian becomes
\begin{equation}
\mathcal{H} = \frac{f_0}{k} \Gi^k a^3 + \pi_a \left(j_0 \Gi^{3k} \right)^{\frac{1}{3}} \;.
\end{equation}
The canonical quantization rules, together with the WDW equation $\mathcal{H} \psi = 0$, provide a system of differential equations of the form
\begin{equation}
\begin{cases}
\displaystyle \pi_{\Gi} \psi = -i \frac{\partial}{\partial \Gi} \psi \;\;\;\;\; \to \;\;\;\;\; \psi(a,\Gi) = A(a) \; \displaystyle \exp\left\{i \; \frac{8 f_0 j_0 \Gi^{4k-1}}{1-4k}\right\}
\\
\mathcal{H} \psi = 0 \;\;\;\; \to \;\;\;\; \displaystyle\frac{f_0}{k} \left(j_0 \right)^{-\frac{1}{3}} a^3 A(a) - i \displaystyle \frac{\partial A(a)}{\partial a} = 0 \;,
\end{cases}
\end{equation}
which can be solved with respect to $A(a)$ to provide the Wave Function of the Universe. The solution is:
\begin{equation}
\psi(a,\Gi) = \psi_0 \exp\left\{i\left[-\frac{f_0}{4k} \left(j_0 \right)^{-\frac{1}{3}} a^4 + \frac{8 f_0 j_0 \Gi^{4k-1}}{1-4k} \right]\right\} \;. \label{WFG}
\end{equation}
Even in this case, the Hartle criterion is preserved by the existence of symmetries, as the wave function is oscillating in the minisuperspace considered. Moreover, in the WKB approximation, the gravitational action can be addressed to the quantity:
\begin{equation}
S = -\frac{f_0}{4k} \left(j_0 \right)^{-\frac{1}{3}} a^4 + \frac{8 f_0 j_0}{1-4k} \Gi^{4k-1},
\end{equation}
so that Hamilton-Jacobi equation with respect to $a(t)$ yields
\begin{equation}
\displaystyle \frac{\partial S}{\partial a} = \pi_a \;\;\;\; \to \;\;\;\; \Gi^k a^3 = 24 \Gi^{k-2} \dot{a}^3 \dot{\Gi} \;,
\end{equation}
which is exactly the first Euler-Lagrange equation. The second Hamilton-Jacobi equation $\displaystyle \left(\frac{\partial S}{\partial \Gi} = \pi_\Gi \right)$ instead, provides the identity $\pi_{\Gi} = j_0$. The system can be implemented with the energy condition, so that the entire system of three differential equations yields the exact solution:
\begin{equation}
a(t) = a_0 t^{1-4k} \;\;\;\;\;\; \Gi(t) = -96 k(1-4k)^3 t^{-4} \equiv \Gi_0 t^{-4} \;.
\end{equation}
As showed in \cite{Bajardi:2020osh}, the  epochs crossed by the universe evolution can be obtained by varying the value of the constant $k$. To conclude, also in this case observable universes are recovered in the semiclassical limit. 

\section{Discussion and Conclusions}
\label{concl}

Minisuperspace Quantum Cosmology   is an approach that ultimately allows to select initial conditions for observable universes. Starting from the Hamiltonian formulation of gravity theories, it is possible, by a quantization procedure,  to obtain the functional WDW equation whose solution is the Wave Function of the Universe. Selecting particular configuration spaces (minisuperspaces), it is possible to reduce the infinite dimensional problem of superspace, and then the WDW equation to a partial differential equation eventually solvable. According to the Hartle criterion, we can determine if dynamical variables of such a wave function are either correlated or not  and then apply the Hamilton-Jacobi equations for  achieving classical trajectories. The  Noether Symmetry Approach, in its Hamiltonian formulation, gives a straightforward interpretation of the Hartle criterion \cite{Capozziello:1999xr,Capozziello:2012hm}: correlations occur for the oscillating components of the Wave Function of the Universe and they are related to the existence of first integrals of motion. The possibility to select classical trajectories (\textit{i.e.}  observable universes) relies on the existence of Noether symmetries. In other words, Noether symmetries constitute a selection rule in Quantum Cosmology.

In this paper,  we considered some classes  of  gravity theories   and selected their functional forms  taking into account  the existence of Noether symmetries.  Specifically, we studied cosmological models related to   $f(R)$  gravity,  $f(T)$ gravity and   Gauss-Bonnet gravity.  
The application of  Noether Symmetry Approach   provides $i)$ the   transformation generator  (which is a symmetry for the starting Lagrangian), $ii)$ the conserved quantity and $iii)$ the form of the action functional. When searching for internal symmetries, it is possible to follow the  procedure  in Eqs. \eqref{relzioni per cambio variabili1} by which the   cyclic variables for the dynamical system are derived. 

After a Legendre transformation,  we obtain the  Hamiltonian function related to the Noether symmetry. After a canonical quantization,  it is possible to derive the corresponding WDW equations and the Wave Function of the Universe. The Hartle criterion is always recovered thanks to presence of first integrals which give rise to oscillating behaviors independently of the considered representation of gravity. This means that classical trajectories, and therefore observable universes, can be always recovered if symmetries exist.
As reported also in \cite{Benetti, Leandros}, the presence of Noether symmetries seems a  criterion to recover physically viable models. 
In a forthcoming paper, this approach will be developed also in comparison with observational data.

\section*{Acknowledgments}

The Authors acknowledge the support of {\it Istituto Nazionale di Fisica Nucleare} (INFN) ({\it iniziative specifiche} GINGER, MOONLIGHT2, QGSKY, and TEONGRAV). This paper is based upon work from COST action CA15117 (CANTATA), COST Action CA16104 (GWverse), and COST action CA18108 (QG-MM), supported by COST (European Cooperation in Science and Technology).

\end{document}